\documentclass[twocolumn,showpacs,preprintnumbers,amsmath,amssymb,pre]{revtex4}
\usepackage{graphicx}
\usepackage{color}

\begin{document}
%\preprint{To be submitted to ?}

\title{Surface tension profiles in vertical soap films}

\author{N. Adami, H. Caps} 
\email{N.Adami@ulg.ac.be}
\affiliation{GRASP -- Universit\'{e} de Li\`{e}ge, physics departement B5, B4000 Li\`{e}ge, Belgium}

\date{\today}

\begin{abstract}
Surface tension profiles in vertical soap films are experimentally investigated. Measurements are performed introducing deformable elastic objets in the films. The shape adopted by those objects  set in the film can be related to the surface tension value at a given vertical position by numerical solving of adapted elasticity equations. We show that the observed dependency of the surface tension versus the vertical position in the soap film can be reproduced by simple modeling taking into account  film thickness measurements.

 \end{abstract}

% 47.15.gm	Thin film flows
% 68.15.+e	liquid film
% 46.25.-y	elasticity in continuum mechanics of solid
% 46.35.+z
% 68.03.Cd	surface tension

\maketitle

\section{Introduction}
\begin{figure}[htbp]
\begin{center}
\includegraphics[width=4cm, height=6cm]{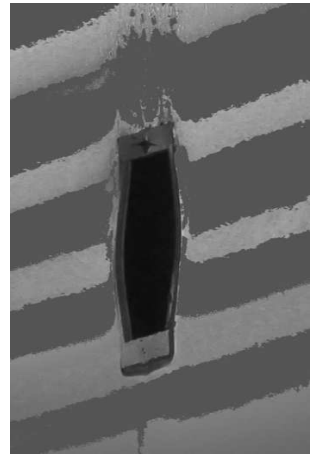}
\caption{Elastic rectangle plugged into a vertical soap film, taken with an angle of 60 degrees regarded to the normal of the film. One can see the deflection experienced by the rectangle due to surface tension forces. Interference fringes are also visible. }
\label{objet_film}
\end{center}
\end{figure}

\noindent Anyone who has ever looked carefully at a soap film just by pulling a frame out of soapy water have seen that it exhibits horizontal interference fringes \cite{Couder1, Nierstrasz, Guyon, Degennes}. As the film has just been pulled out of soapy water, all the fringes seems to be vertically equally separated. After a few seconds, the interfringe rises from the bottom to the top of the film. This fact attests of both the existence of a non-uniform thickness profile in the soap film, and also of the evolution of that profile with time. From a mechanical point of view, this thickness profiles implies that i) for any value of the height above the bottom edge of the frame $H$, the weight of the part of the film which lies beneath this value must be conterbalanced by local surface tension force and ii) since the fringe pattern in the film stretches from bottom to top, the sustaining surface tension profile must increase and saturate with $H$. \\
Due to their particular geometric properties, soap films have been in the center of numerous studies these last decades \cite{Mysels, Couder1, Kellay1, Bruinsma, Nierstrasz, Scheid1, Wu2, Rivera1, Rutgers, Zhang_wu, Prasad}. Early works by Mysels et. al.  \cite{Mysels, Prasad, Schwartz} showed that their behaviours strongly depend on parameters such as viscosity \cite{Prasad}, but also on the chemicals used to produce the films. Specific surfactant-linked phenomena such as marginal regeneration \cite{Mysels, Nierstrasz} have also been spotted to account dramatically in phenomena such as drainage, leading to unexpected film lifetimes. Those studies of isolated soap fillms have been widely used to model global behaviours of more complex system such as foams \cite{Weaire}. Theoretical expressions based on surfactant molecules thermodynamics have been proposed in order to describe the evolution of surface tension versus $H$ \cite{Couder1}. To our knowledge, no experiment has yet been proposed in order to confirm those theoretical trends.\\
Recent studies on soft elastic objects have shown that considerable deformations of those objects can be induced by capillary constraints \cite{Landau, Zell, Hu, Choidi, Py, Py_2, Benoit, Jose, Adami2}. The case of surface pressure linked to a surfactant monolayer leading to deformations  \cite{Degennes, Guyon, Zell, Hu, Adami2} can either be used to characterize the evolution of the object shapes knowing the surface pressure, or to determine the surface pressure leading to an observed deformation.\\
In this paper, we show how elasto-capillary effects can be used in order to probe surface tension profiles of  vertical maintained soap films. Figure \ref{objet_film} shows a soft rectangle which has been plugged into a vertical soap film. When no soap film is present inside the rectangle, this latter inflates due to surface tension forces applied by the film interfaces. Solving the elasticity beam equations, written in this particular case allows to get a theoretical link between the deflections experienced by the objects and the surface tension value at a given vertical position. The experimental surface tension profiles can be reproduced using a combination of a simple model for the mechanical equilibrium in the film and thickness measurements in the film.

\section{Experimental setup and materials}
\subsection{Soap films}
\begin{figure}[htbp]
\begin{center}
\includegraphics[height=4cm, width=6cm]{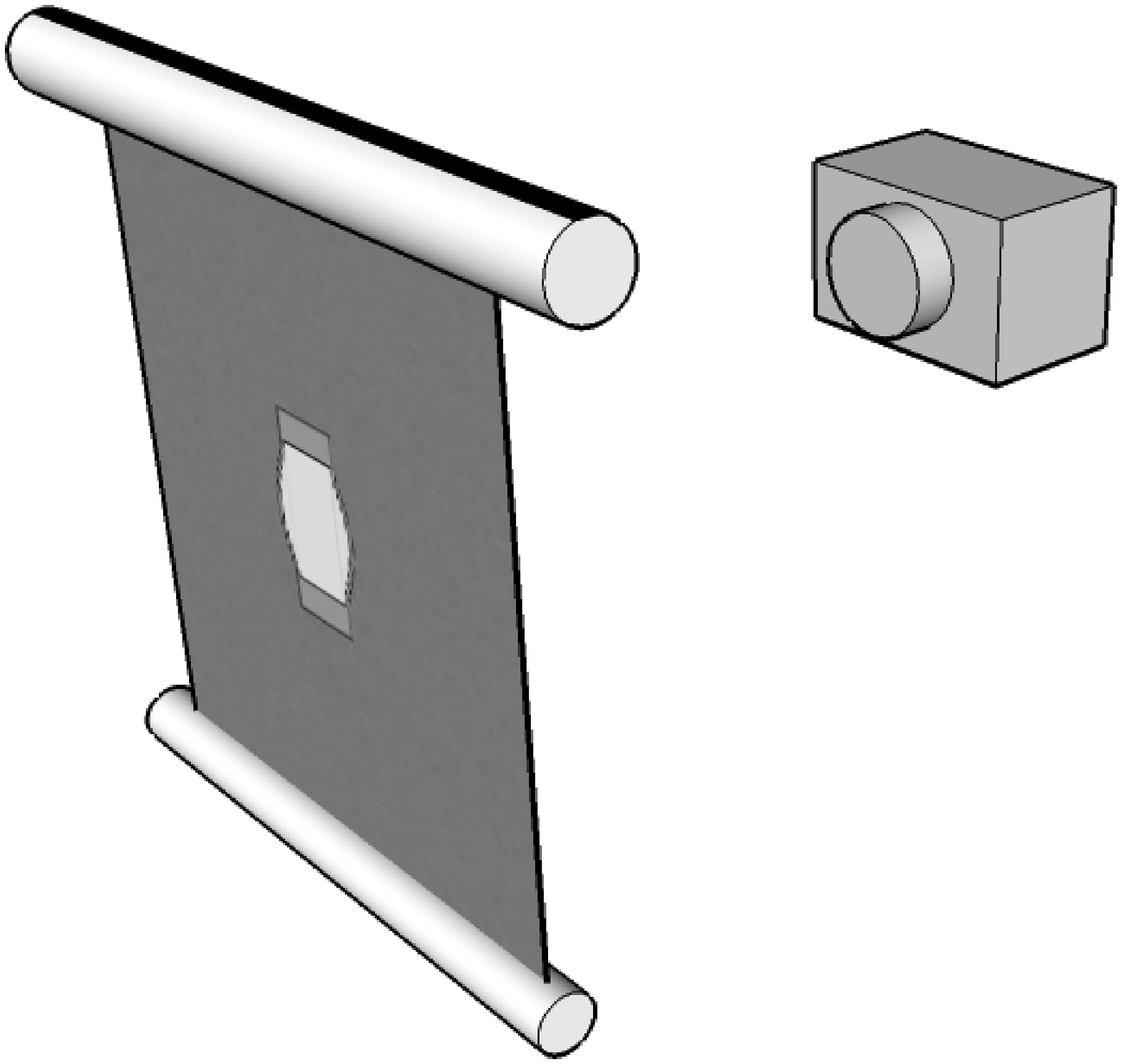}\ \includegraphics[height=4cm, width=2.5cm]{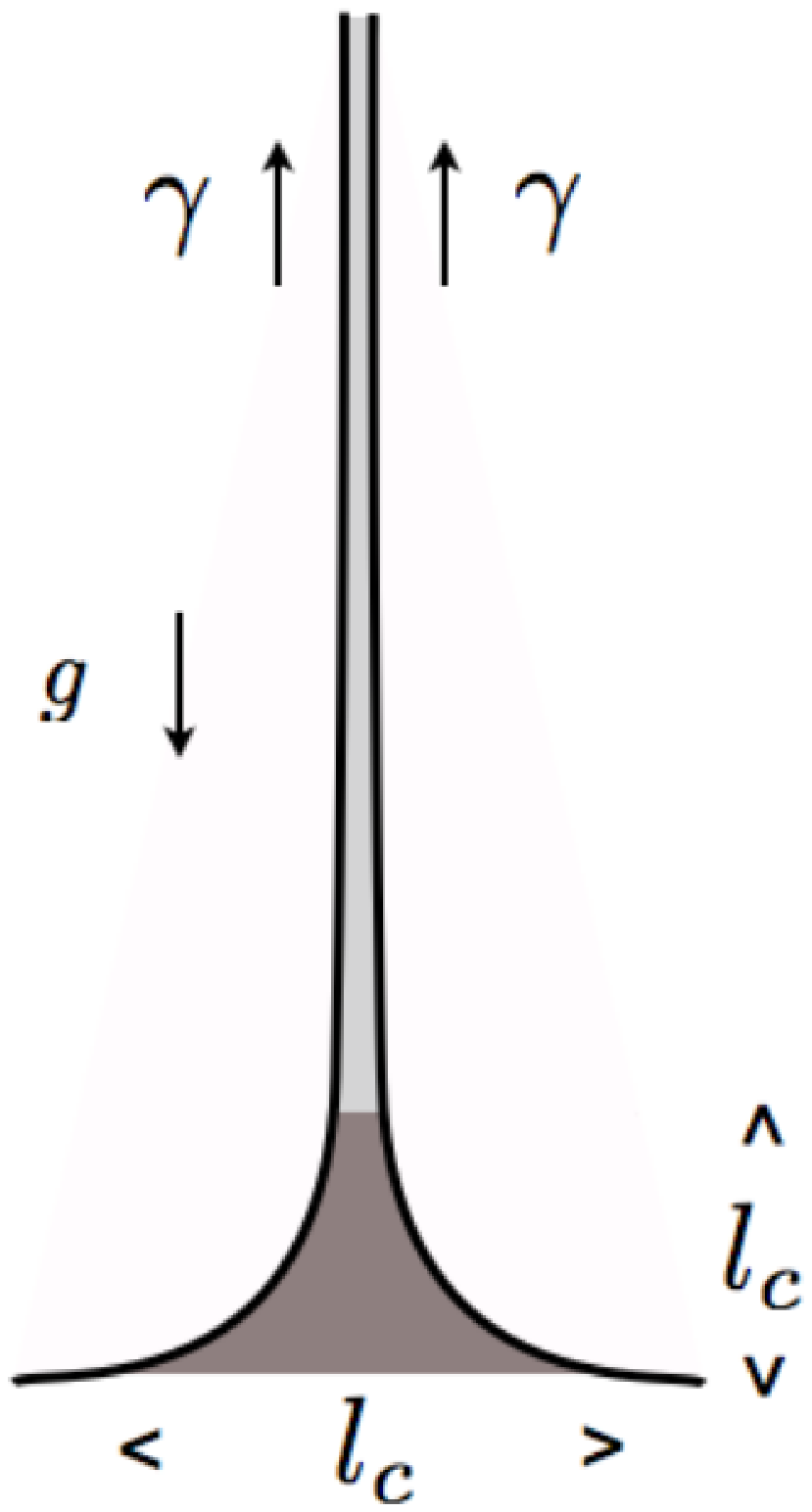}
\caption{Left : Sketch of the experimental setup. The slit pipe, the soap film, the camera and the deflected object are represented. Right : scheme of bottom meniscus linking the film to the bottom of the frame (see next section for details).}
\label{setup}
\end{center}
\end{figure}

Soap films are built from solutions made of 3 \% of a solution of SLES+CAPB described in \cite{Denkov} plus 0.3 \% glycerol and double-distilled water. This mixture leads to typical density $\rho=1000$ $\rm{kg/m^3}$ and  viscosity $\eta=\rm{10^{-3}}$ Pa.s. The surface tension $\gamma_0$ of the solution is determined to be $29.8 \pm 0.2$ mN/m. In order to avoid the temporal evolution of the film, we fix their thickness profile by feeding them thanks to a setup sketched at Fig.\ref{setup} \cite{Brunet1, Brunet2, Adami1}. A flow made of the soapy solution is injected with a constant flow rate $Q$ by both sides in a slit pipe, the slot pointing upward. When soapy water gets out of the slot, it follows the edges of the pipe to reach the top of the film which lies beneath the pipe. Doing so, it is possible to suppress the thinning of the film. Choosing the flow rate carefully allows to fix the thickness profile of the film. The thickness dependency versus $H$ has been determined in  \cite{Adami1} . Such measurement reveals that the thickness profiles of our films scale like power laws, say $e(H)=aH^{-\beta}$ with $a$ being a scaling constant and $\beta$ the scaling exponent.\\
 The lateral and bottom edges of the frame sustaining the film are made of stainless steel rod. All the edges of the film present a length $L=150$ mm, fixed for all experiments described below.\\ %Constant flows are created thanks to a constant flow pump.\\ %The $Q$ values used in this study are 1.8, 2.0 and 2.2 ml/min.\\

\subsection{Elasto-capillary probe}

\begin{figure}[htbp]
\begin{center}
\includegraphics[width=6cm, height=4cm]{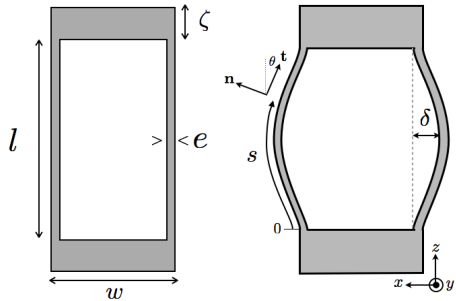}
\caption{Left : Geometrical feature of the rectangles used to perform surface tension measurements in soap films ($l=24 \rm{mm}$, $w=12\rm{mm}$, $\zeta=3\rm{mm}$, $e=0.8\rm{mm}$). Right : typical shape adopted by rectangles when introduced in the soap films.}
\label{rect}
\end{center}
\end{figure}

The elastic objects used to perform surface tension measurements are vinylpolyxilosane rectangles presented on Fig.\ref{rect} \cite{Zell, Hu}. They are made of long a flexible lateral arms of length $L$ and rigid top and bottom arms of height $\zeta$, the global width of the rectangle being $w$.The section of the lateral arm is square with size $e$. As presented on Fig.\ref{setup} the probe is maintained in the film at a given height thanks to a needle. Since no soap film is present inside the rectangle, surface tension forces acting on the arms will cause the bending of the lateral arms, inflating the rectangle. This situation is analog to the one described in \cite{Zell, Hu} . The amplitude of the corresponding deflection $\delta$ results in a balance between interfacial and bending energies. The bending energy $B$ of a flexible arm is related to its quadratic momentum as $EI_0$, where $E$ is the Young Modulus of the polymer and $I_0=e\zeta^3/12$  its quadratic momentum. In the particular case of the lateral arm, this expression reduces to $e^4/12$. The width $\zeta$ is chosen  so that only the lateral edges can bend once the object is set in the film. The corresponding Young modulus and density of the rectangles are respectively 0.3 MPa and 1023 kg/$\rm{m^3}$.\\

\begin{figure}[htbp]
\begin{center}
\includegraphics[height=4cm, width=2.5cm]{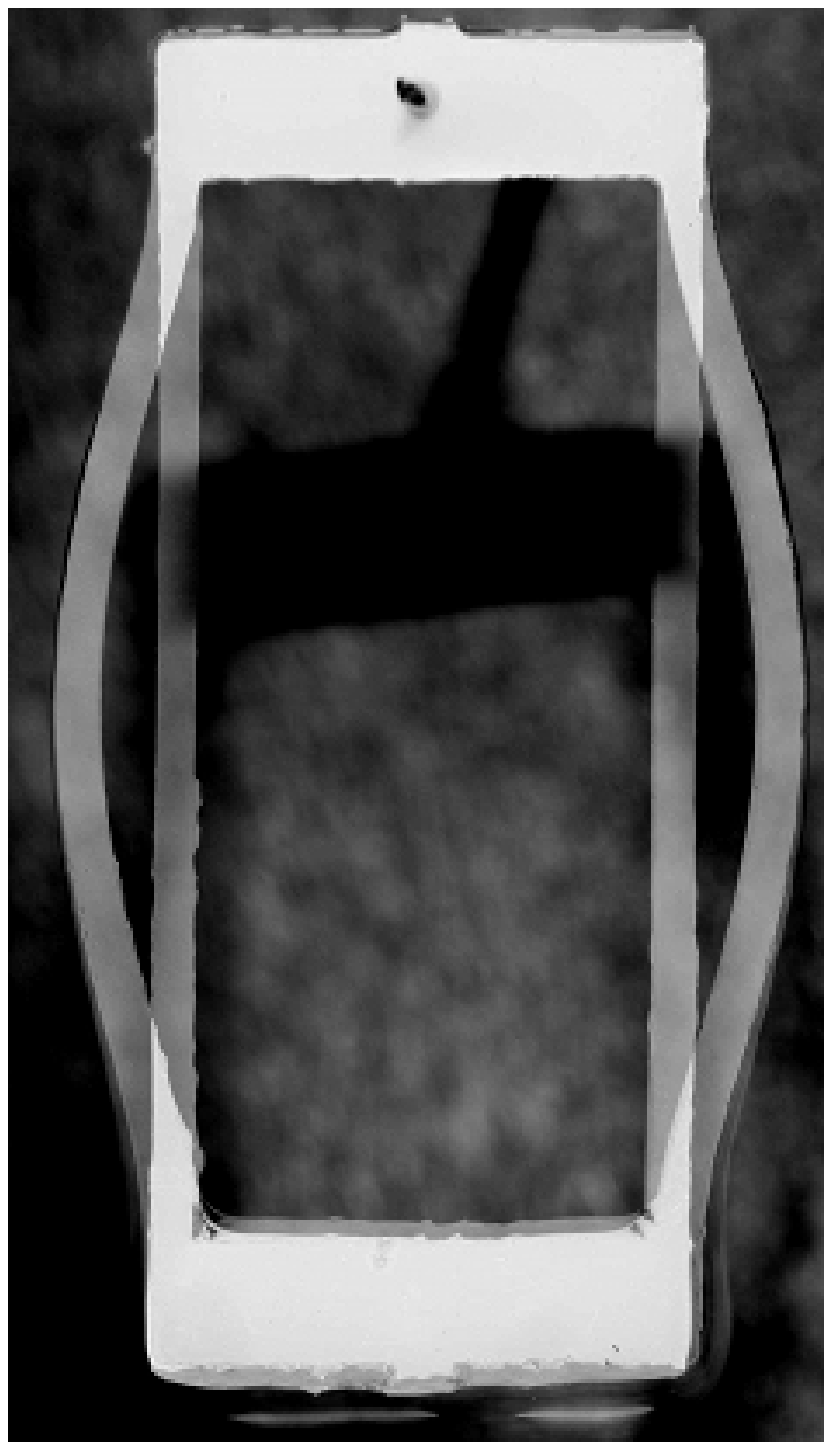}\ \includegraphics[height=4cm, width=6cm]{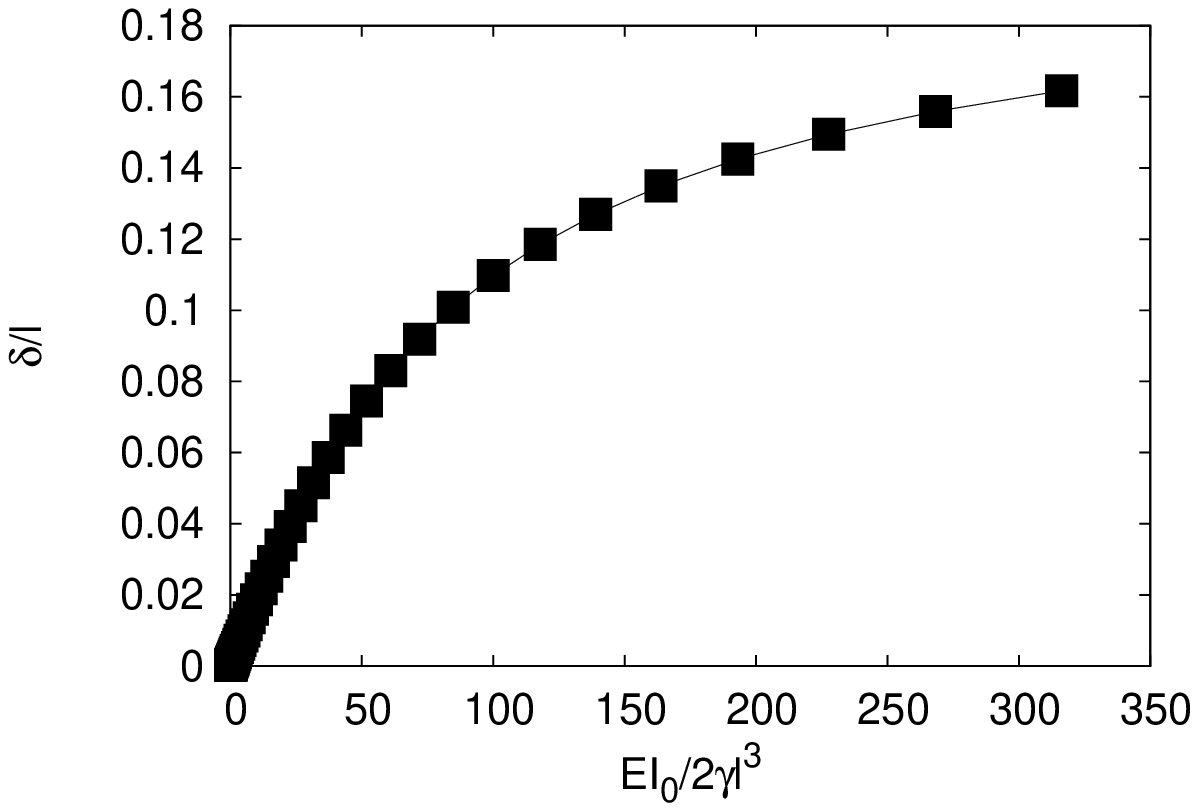}
\caption{Left : superposition of a rectangle when set inside and outside the soap film. Both $\delta$ and $l'$ can be seen on this picture. Right : adimensionned deflection $\delta$ versus the control parameter $EI_0/2\gamma l^3$, as obtained from Eq.\ref{force_interne} and Eq.\ref{elastica}. }
\label{superposition_calib}
\end{center}
\end{figure}

It is instructive to estimate the different quantities involved in our experiment by order of magnitude calculations. Figure \ref{superposition_calib} shows a superposition of pictures of the rectangle in the soap film when soap film is present both inside and outside the rectangle (undeformed) and when the inner soap film is burst (inflated case). A careful look at the bottom edge of the rectangle allow to see that it rises up of a quantity $l'$ due to the film stretching. This effect involves an increase of potential energy of both the bottom part of the rectangle and of the meniscus linking this arm to the film. Including this potential energy the total energy balance after the inner film burst reads : 
\begin{equation}\label{nrj_film}
2 \gamma\delta l\sim EI_0\Big(\frac{\delta}{l^2}\Big)^2l+\gamma w l'+\rho_SV_Sgl'+\rho_LV_Lgl'
\end{equation}
\noindent where the first term of the right member is the bending energy linked to $\delta$,  $\gamma w l'$ the loss of interfacial energy due to the rise of the bottom part, $\rho_SV_Sgl'$ the increase of potential energy of the bottom part of the rectangle, $\rho_S$ and $V_S$ being the density and volume of that part, and $\rho_LV_Lgl'$ the increase of potential energy for the meniscus. As one can see on Fig. \ref{superposition_calib}, the inner film burst leads to $l'$ which are weak compared to $\delta$, this later being relatively weak compared to $l$. Thus, from a geometrical point of view, $l'$ can be estimated as $\delta ^2/l$. Equation (\ref{nrj_film}) then reads : 

\begin{equation}
2 \gamma\delta l\sim \frac{\delta ^2}{l}\Big(\frac{EI_0}{l^2}+\gamma w+\rho_s V_S g +\rho_LV_Lg\Big)
\end{equation}

\noindent Introducing order of magnitude in this formula leads to an estimate for the surface tension of $\gamma\sim10^{-2}$ N/m, which is consistent with typical values of surface tension. We can then estimate the fluctuation of the typical deflection encountered by the lateral arm. From thickness measurement (see \cite{Adami1}) mechanical equilibrium in the film, we can estimate the surface tension variation to be $\sim 0.2$ mN/m if $H$ is increased/decreased by 1cm. We can also spot from Eq. (\ref{nrj_film}) that the bending energy is weaker than the other terms. We can then assume that a change in $H$ mainly leads to a change in $\delta$ rather than in $l'$, which is in agreement with the geometrical expression of $l'$ as a function of $\delta$. Increasing $H$ will lead to $\delta$ fluctuations which are weak regarded to the typical $\delta$ value linked to the inner film burst. Since the curvature linked to the lateral arm is expressed as $\delta /l^2$ and potential energy increments are proportional to $\delta^2/l$, it seems logical to consider that increasing $H$ only tend to create curvature rather than elevation of the lower part of the frame. We can thus exclusively consider bending energies to estimate the corresponding deflection as $\Delta\delta\sim\Delta\gamma l^4/16 eE^4$, to be close to 2e-5 m. This estimation implies that both measurements and rectangles have to be designed so that the resolution of the setup is high enough to see the variations of $\delta$ with $H$.\\

While the previous calculation gives consistent estimations  for $\gamma$, further formalization is needed in order to get the accurate dependency of $\delta$ versus $\gamma$. The typical beam equations governing the evolution of the shape of a deformable object due to an external constraint are \cite{Landau} : 

\begin{equation}\label{force_interne}
\frac{d\mathbf{F}}{ds}=-\mathbf{K}
\end{equation}

\begin{equation}\label{elastica}
EI_0\frac{d^2 \theta}{ds^2}=(\mathbf{F} \times \mathbf{t})\cdot \mathbf{e_y}
\end{equation}
\noindent where Eq.(\ref{force_interne}) is the relationship between the internal forces of the object $\mathbf{F}$ and the external constraint per unit length $\mathbf{K}$, while Eq.(\ref{elastica}) is the well-known Euler's Elastica, expressing the torque equilibrium in the object. This system is expressed in the curved coordinate $(s,\theta)$ as represented on Fig.\ref{rect}. For symmetry reasons, these equation are solved for $s\in [0,l/2]$. Assorted with its boundary condition, $\mathbf{K}$ reads :  

\begin{equation}\label{K}
\mathbf{K}=2\gamma\mathbf{n}-\rho_S g A \mathbf{e_z}
\end{equation}
\noindent  with $A$ is the section of the lateral arm. This expression can then be included in the system formed by Eq.(\ref{force_interne}) and Eq.(\ref{elastica}). These equation can then be adimensionned for easier computational investigation, leading to the appearance of $EI_0/2\gamma l^3$ as control parameter. This quantity can also be deduced from simple order of magnitude calculation, and appears in many elasto-capillary driven systems \cite{Landau, Zell, Hu, Choidi, Py, Py_2, Benoit, Jose}. To be complete, we have to mention that a second control parameter exist in the system, being $EI_0/\rho g S_0$. For the purpose of our measurements, the physical and geometric properties of the rectangles are fixed as described previously, so that the latter parameter is fixed. \\
As previously emphasized, it is necessary to include the downward pulling of both gravity and surface tension on the bottom part of the rectangle. In order to take them into account in the numerical solving, the initial condition on the force along $\mathbf{e_z}$ must be equal to $-\rho_S V_Sg-\rho_L V_L g-\gamma w$, which has to be expressed as a function of the control parameter. The other boundary condition used are $\theta(s=0)=0$, $\theta(s=l/2)=0$ and $F_x(s=l/2)=2\gamma e$. This system is then numerically solved using a fourth order Runge-Kutta method. Since $\delta$ is the easiest quantity to measure in experiments, we decided to consider the evolution of the adimensioned deflection $\delta/L$ as a function of the solving parameter (Fig.\ref{superposition_calib}). This curve is later used to link the deflections obtained from the experimental data to the value of the corresponding constraint. Doing so,  it is possible to determine the evolution of surface tension with the vertical coordinate in the soap films. The deflection as a function of heigh was recorded with an optical resolution of 4.6e-6 m, leading to a typical accuracy of 20\% for $\delta$ measurements.

\section{Theory and modeling}
%We here discuss the different theoretical ideas involved in this work. We first describe how the mechanical equilibrium has to be considered in order to describe properly how the surface tension is expected to be behave, taking into account the different features of the films. We then describe how beam elasticity equation can be used to predict how the shape of rectangles describe above are supposed to evolve with the vertical coordinate.

\subsection{Soap film mechanical equilibrium}
 The existence of a soap film implies that for any value of the $H$, surface tension forces must carry the weight of the liquid lying under $H$. Careful observation of the film a few second after its formation show that the bottom meniscus linking the film to the bottom of the frame is always greatly thicker than the lateral and top ones, due to drainage phenomena. This meniscus should then be taken into account when estimating the total weight to be balanced by surface tension. The right part of Fig.\ref{setup} presents a scheme of the transverse view of the situation at the bottom of the film (not to scale). Basically, the soap film must be considered as made of a meniscus plus a thinner part, the last one being the only one concerned with surface tension measurements, the typical height of the meniscus being smaller than $L$.\\ 
In the following calculation, we consider that the thickness profile of the soap film does not evolve in time, thanks to the feeding. Considering a given value $H$ of the vertical coordinate, the mechanical equilibrium writes : 
\begin{equation}\label{equilibre_dynamique}
2\gamma(H)L= W_{film}+W_{meniscus}
\end{equation}
where $W_{meniscus}$ can be expressed as : 
\begin{equation}\label{w_menisque}
W_{meniscus}=\rho l_c^2Lg=\gamma_0L
\end{equation}
\noindent with $l_c=\sqrt{\gamma_0/\rho g}$ is the capillary length. In fact, it seems logical to consider that the surface tension does not exhibit strong variations in the meniscus and must then be close to the value of a planar interface, say $\gamma_0$. 
The weight of the thin part of the film must be linked to the thickness profile of the film as :

\begin{equation}\label{W_film}
W_{film}=\rho g L \int_{0}^{H} e(H) dH
\end{equation}
\noindent with $e(H)$ the thickness profile of the film. One sees that the integral in Eq.(\ref{W_film}) must both increase and saturate with $H$, since the film is thicker at the bottom.

%\subsection{Elasto-capillary sensor}

 \section{Measurements}
\begin{figure}[htbp]
\begin{center}
\includegraphics[height=6.5cm, width=9cm]{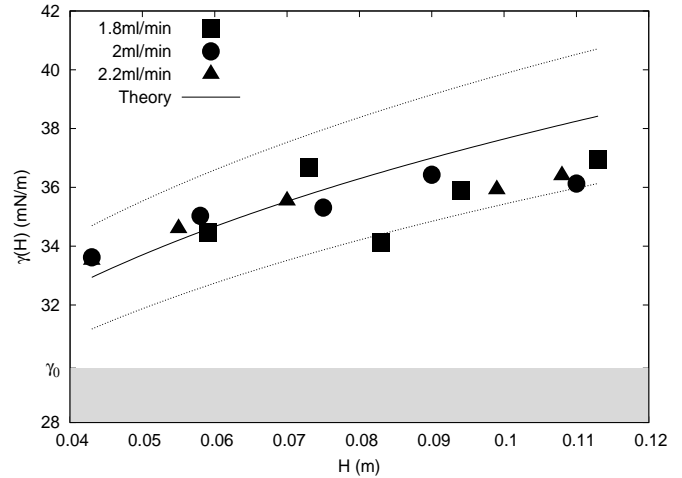}
\caption{Surface tension profiles in vertical soap films as a function of the vertical position. The solid line is expression (\ref{gamma_versus_H}) expressed for the mean $a$ value, while dashed lines show the dispersion of this expression around the mean curve. Values of $\gamma<\gamma_0$ are not physically permitted and correspond to the grayed area.}
\label{profil_gamma}
\end{center}
\end{figure}

The calibration curve presented on Figure \ref{superposition_calib} relates the experimental deflection $\delta$ to the corresponding value of $\gamma$. Since the thickness profile is known \cite{Adami1}, the analytic expression for the surface tension profile as a function of $H$ obtained by combination of Eq. (\ref{equilibre_dynamique}), Eq. (\ref{w_menisque}) and Eq. (\ref{W_film}) reads : 

\begin{equation}\label{gamma_versus_H}
\gamma(H)=\frac{\gamma_0}{2}+\frac{\rho_L g a}{2(1-\beta)}H^{1-\beta}
\end{equation}

where all the quantities are known from experimental determinations. Figure \ref{profil_gamma} shows the superposition of experimental values of $\gamma(H)$ for different $Q$ values and (\ref{gamma_versus_H}). The agreement between experimental values and theoretical predictions attests from the validity of our simple model to predict the evolution of surface tension in vertical soap films.

\section{Discussion}
Despite agreement between experimental data and theory, one could wonder why points corresponding to different $Q$ values are mixed rather than being separated on different curves. In fact, since an increase in $Q$ is supposed to lead to an increase in $e(H)$ (see \cite{Adami1}), the weight of the film, and so the surface tension profiles are supposed to increase as well. In order to get reliable measurements, we measured $\delta$ several times for each value of $H$ and $Q$, and took the averages, shown on Fig.\ref{profil_gamma}. Those average points mix all together due to inherent fluctuations of the system (see \cite{Adami1}). Moreover, as previously emphasized, the error on $\delta$ measurement is close to 20\%, and the $\delta$ measurement procedure implies the Young modulus $E$, which value is known with a relative precision of 10\%. Nevertheless, the most important result is that measurements presented on Figure \ref{profil_gamma} show that the evolution of the surface tension with $H$ does exist in vertical soap films, and that this evolution can be described faithfully by taking into account the mechanical equilibrium in the film. Despite this trend has been predicted by Gibbs in the late 1880, this is the first time, to our knowledge, that it is experimentally evidenced. 

\section{Conclusion}
In conclusion, we have investigated surface tension profiles in maintained soap films. We designed an elasto-capillary sensor to probe those surface tension profiles in the film. We solved typical elasticity equation in order to link experimental rectangle deflections with the corresponding surface tension values. Repeating the experiment several times for several vertical position allowed us to get mean surface tension profiles. We were able to propose a simple model which brought that both the general trend got from the surface tension measurement method here presented and previous thickness measurements are in good quantitative agreement.\\
A perspective to this work would be to perform the same experiment with other surfactant, and to couple them to thickness measurements in order to check the relevance of our assumptions to describe soap films.\\

\noindent\textbf{Acknowledgements} : This work was financially supported by FRS-FNRS. The authors would like to thank J. Bico, B. Roman and J. Seiwert for helpful discussions.

%\Large{Ce qui reste à faire dans ce papier : Inclure la biblio dans le texte (done)+rajout éventuel\\
%Pacs (done)\\
%faire relire\\
%eventuellement une photo des objets dans le film et sans film, pour montrer l'augmentation d'énergie potentielle de la partie inférieure.\\
%refaire des mesures pour les parties hautes des films, pour eventuellement améliorer l'accord\\
%faire une mesure correcte de $\gamma_0$\\
%...}

%\newpage

\end{document}